# A Compact xDAWN-Guided Spatio-Temporal Network with Filter-Bank Convolution for P300 Detection

**Mohammad Mahdi Habibi Bina**[1], **Mohammad Reza Daliri**[1]
[1] Neuroscience & Neuroengineering Research Laboratory, Biomedical Engineering Department, School of Electrical Engineering, Iran University of Science and Technology (IUST), Tehran, Iran

**ABSTRACT**
P300-based brain–computer interfaces (BCIs) enable communication by decoding stimulus-evoked neural responses into character selections without relying on voluntary motor output. However, reliable P300 decoding remains challenging because of the low amplitude and inherently low signal-to-noise ratio (SNR) of the response, while practical P300 spellers require accurate character selection with a limited number of stimulus repetitions and low computational complexity. We propose a compact, deep learning architecture that integrates established EEG signal-processing priors with end-to-end representation learning for P300 detection. The proposed architecture combines xDAWN spatial filtering for ERP enhancement and dimensionality reduction with a fixed frequency-selective filter bank and convolutional feature learning to extract discriminative spatio-temporal representations directly from EEG. Depthwise-separable convolutions and progressive average pooling further reduce the computational complexity while preserving task-relevant temporal information, resulting in an end-to-end target/non-target decoder with only 5,100 trainable parameters. The proposed method was evaluated on two publicly available P300 speller benchmarks, BCI Competition II—Dataset IIb and BCI Competition III—Dataset II, and compared with state-of-the-art P300 decoding methods. Across different numbers of stimulus repetitions, the proposed architecture achieved a favorable accuracy–speed trade-off. On BCI Competition III, it achieved an information transfer rate (ITR) of 17.8 bits/min with only two repetitions, while achieving an ITR of 40.5 bits/min on BCI Competition II with consistently strong character-selection performance. These results demonstrate that incorporating domain-informed spatial and spectral priors into a compact deep-learning architecture can improve P300 decoding efficiency while substantially reducing model complexity, supporting fast and computationally efficient EEG-based BCI communication.
**INDEX TERMS** Neural Decoding, Computational Neuroscience, Brain-Computer Interface (BCI), EEG, P300

## 1. INTRODUCTION

Brain–computer interfaces (BCIs) provide a direct communication pathway between the brain and external devices by translating neural activity into actionable commands [1]. Based on the recording modality, BCIs can be broadly categorized as invasive or non-invasive systems. Among non-invasive approaches, electroencephalography (EEG) is particularly attractive because it is relatively safe, inexpensive, and capable of providing high temporal-resolution measurements [2], despite its inherently low signal-to-noise ratio (SNR) [3],[4]. Several EEG components can support BCI control, including sensorimotor rhythms, steady-state visual evoked potentials, and event-related potentials (ERPs). ERPs are time-locked neural responses elicited by specific sensory, cognitive, or motor events and have been extensively exploited in BCI systems [5]. Among them, the P300 is a prominent ERP component elicited by infrequent target stimuli within an oddball paradigm, typically emerging approximately 300–600 ms after stimulus onset [6], [7], [8]. In visual P300 spellers, differences between target and non-target responses are exploited to identify the row and column containing the attended character, enabling character-by-character communication without conventional motor output [7]. The P300 speller paradigm introduced by Farwell and Donchin [9] established a foundation for EEG-based character selection and remains one of the most widely studied ERP-based BCI paradigms.

Early P300 decoding approaches primarily relied on conventional machine-learning pipelines combining signal preprocessing, feature extraction, and statistical classification. For example, Rakotomamonjy et al. [10] achieved 96.5% classification accuracy using recursive channel elimination and an ensemble of support vector machines (ESVM) over 15 repetitions. Such approaches demonstrated that careful feature and channel selection could substantially improve P300 discrimination, particularly under the pronounced target/non-target class imbalance inherent to the oddball paradigm. Beyond classification, spatial filtering methods such as xDAWN have provided an effective means of enhancing the weak ERP response by exploiting its reproducible spatial and temporal structure [11]. xDAWN has subsequently been combined with classifiers including MLP, BLDA, and linear SVM, yielding improved P300 decoding performance [12]. Its effectiveness has also been demonstrated in competitive P300 decoding settings through combinations with channel selection, regularization, and Riemannian methods [13], [14], [15], [16]. Collectively, these studies highlight the value of explicitly incorporating the spatial and signal-processing structure of ERP signals into P300 decoding.

More recently, deep learning has shifted P300 decoding from manually engineered features toward end-to-end representation learning. Most successful P300 decoders have adopted convolutional architectures, including EEGNet [17], DeepConvNet [18], ShallowConvNet [18], CNN-based models [19], [20], and BN3[21], which learn discriminative spatial and temporal EEG representations directly from short EEG epochs. Given that a typical P300 trial spans only 600–800 ms, temporal convolutions efficiently capture the local ERP dynamics while maintaining a compact parameter footprint, making CNNs the dominant architecture in this domain.

Subsequent advances explored richer representation-learning mechanisms beyond conventional convolutions. ERP-CapsNet introduced the first capsule network for ERP detection, demonstrating that capsule representations could better preserve the hierarchical characteristics of P300 responses than conventional CNNs [22]. Building on this idea, ST-CapsNet incorporated spatial and temporal attention modules before capsule-based feature extraction, enabling adaptive refinement of task-relevant EEG features and further improving P300 detection performance [23].

Despite these advances, several fundamental challenges remain in practical P300 speller systems. The P300 is a low-amplitude ERP embedded in noisy multichannel EEG, making reliable detection difficult because of the inherently low signal-to-noise ratio of EEG recordings. Moreover, the Oddball paradigm produces a pronounced class imbalance between target and non-target trials, which complicates model optimization. Beyond accurate trial classification, practical P300 spellers must also address the speed–accuracy trade-off: increasing the number of stimulus repetitions generally improves character recognition accuracy but proportionally increases the time required for character selection. Consequently, Information Transfer Rate (ITR), which jointly quantifies decoding accuracy and communication speed, has become the primary metric for evaluating practical P300 speller performance.

In this study, we propose a compact decoder designed to improve communication efficiency while preserving reliable P300 detection. The proposed architecture combines xDAWN-based ERP enhancement, fixed FIR filter-bank decomposition into six physiologically meaningful EEG frequency bands, and depthwise spatio-temporal representation learning, while employing Focal Loss to mitigate the inherent class imbalance of the Oddball paradigm. Together, these design choices enable discriminative P300 representations to be learned from fewer stimulus repetitions while maintaining a lightweight architecture suitable for efficient EEG decoding.

The description of our contributions in this research:

1. We propose a compact xDAWN-guided spatio-temporal network with fixed filter-bank convolution for P300 detection, integrating ERP-oriented spatial enhancement, six physiologically meaningful EEG frequency bands, and depthwise spatio-temporal representation learning within an end-to-end framework.
2. We incorporate xDAWN spatial filtering as an ERP-oriented prior to enhance P300-related activity under low-SNR EEG recordings before hierarchical feature learning.

3. We address the inherent target/non-target class imbalance of the P300 speller paradigm by employing Focal Loss to improve the learning of difficult samples during training.
4. We demonstrate that the proposed method achieves state-of-the-art performance on the BCI Competition II (Dataset IIb) and BCI Competition III (Dataset II, Subjects A and B), while maintaining a compact architecture with only 5,100 trainable parameters and superior information transfer rate at low repetition numbers.

The remainder of this paper is organized as follows. Section 2 describes the datasets and preprocessing procedures, followed by a detailed presentation of the proposed method and its components. Section 3 presents the experimental results across different datasets and compares the proposed method with representative state-of-the-art approaches. Finally, Section 4 concludes the paper.

## 2. MATERIALS AND METHOD DETAILS

### *2.1. Data and Preprocessing*

We evaluated the proposed decoder on two publicly available (The dataset is available at https://www.bbci.de/competition/) and widely used P300 speller benchmarks from the BCI Competition: BCI Competition II—Dataset IIb and BCI Competition III—Dataset II. As can be seen in Figure 1, both datasets were acquired using the Farwell–Donchin row/column P300 speller paradigm, in which the user attends to a target character within a matrix while rows and columns are sequentially intensified. The datasets therefore provide a consistent experimental framework for evaluating single-trial target/non-target discrimination and downstream character selection. The official BCI Competition records identify Dataset IIb as a 36-class P300 speller dataset with 64 EEG channels sampled at 240 Hz, while Dataset II of BCI Competition III contains recordings from two subjects under the same P300 speller paradigm.

BCI Competition II—Dataset IIb contains recordings from a single subject across three sessions. Sessions 10 and 11 were used for training and Session 12 for evaluation. The training data comprise 42 characters, each presented over 15 repetitions, with two target and ten non-target stimuli per repetition, resulting in 1,260 P300 ($42 \times 2 \times 15$, corresponding to 42 characters × 2 target flashes per repetition × 15 repetitions) and 6,300 non-P300 trials ($42 \times 10 \times 15$). The evaluation set contains 31 characters and consequently 930 P300 and 4,650 non-P300 trials. This yields the characteristic target-to-non-target imbalance of the row/column P300 paradigm.

BCI Competition III—Dataset II contains recordings from two subjects (Subject A and Subject B). For each subject, the training set comprises 85 characters, each presented over 15 repetitions, yielding 2,550 P300 and 12,750 non-P300 trials. The corresponding test set contains 100 characters and 3,000 P300 and 15,000 non-P300 trials.

For both datasets, the continuous EEG recordings were band-pass filtered between 0.1 and 50 Hz and subsequently segmented into 800-ms post-stimulus epochs, resulting in 192 temporal samples per trial at 240 Hz. Each epoch was assigned a binary label corresponding to a target (P300) or non-target (non-P300) stimulus. The resulting datasets preserve the substantial class imbalance inherent to the original paradigm while providing a standardized input representation for the proposed network.

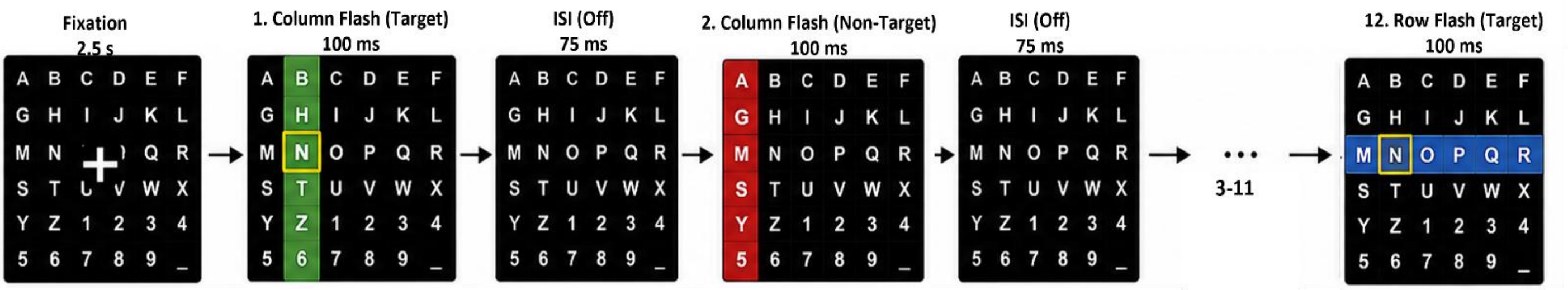

*Figure 1. Visual P300 speller paradigm based on the Farwell–Donchin 6 × 6 row–column matrix. A sequence consists of 12 randomized stimulus intensifications, including six row and six column flashes, with a flash duration of 100 ms and an inter-stimulus interval (ISI) of 75 ms. For the target character "N", the second-column and third-row flashes constitute the target (P300) stimuli, whereas the remaining ten intensifications are non-target stimuli. The 2.5-s fixation period precedes each stimulus sequence.*

The complete distribution of target and non-target trials across the training and evaluation sets is summarized in Table 1.

*Table1 . Number of P300 and Non-P300 Trials in the Training and Test Datasets.*

| **Dataset** | **Train** | | **Test** | |
|---|---|---|---|---|
| | **P300** | **Non-P300** | **P300** | **Non-P300** |
| BCI competition II-Dataset IIb | 1260 | 6300 | 930 | 4650 |
| BCI competition III-Dataset II (Subject A) | 2550 | 12750 | 3000 | 15000 |
| BCI competition III-Dataset II (Subject B) | 2550 | 12750 | 3000 | 15000 |

### *2.2. Proposed Architecture*

The proposed architecture is illustrated in Figure. 2. We developed a compact, neurophysiology-informed deep learning decoder that integrates established EEG signal-processing priors with hierarchical spatio-temporal representation learning for P300 detection. The individual components and processing stages of the architecture are described in the following subsections. Given short multichannel EEG epochs, the proposed architecture first performs xDAWN-based spatial projection to enhance ERP-related activity and reduce the spatial dimensionality, followed by a fixed FIR filter bank that explicitly decomposes the signal into six physiologically relevant frequency bands. Frequency-specific spatial representations are then learned through depthwise convolutions, while depthwise-separable temporal convolutions capture discriminative temporal dynamics with a substantially reduced parameter footprint. Progressive temporal downsampling further yields a compact representation that is ultimately mapped to a binary target/non-target decision. This design preserves interpretable spectral and spatial priors while replacing handcrafted feature extraction with learnable representations, providing a unified framework that jointly addresses P300 discriminability, computational efficiency, and rapid decoding. The following subsections detail the input representation and the successive operations of the proposed network.

**Input Layer:** The input data are represented as a tensor with dimensions ((N, T, C)), where (N) denotes the number of EEG trials, (T) represents the number of temporal samples per trial, and (C) denotes the number of EEG channels. Each trial contains 64 EEG channels sampled at 240 Hz over an 800-ms temporal window, resulting in (T=192) temporal samples. For example, for Subject A of the BCI Competition III dataset, the training set contains 2,550 trials, yielding an input tensor of dimensions ((2550, 192, 64)). An additional singleton dimension is subsequently introduced to accommodate the two-dimensional convolutional operations employed by the network.

**Spatial Filtering Layer (xDAWN):** The low signal-to-noise ratio (SNR) of EEG recordings makes P300 detection inherently challenging because the recorded signals contain both task-related neural activity and substantial background activity originating from unrelated brain processes. Moreover, P300 responses are not distributed uniformly across electrodes, with parietal channels generally carrying more discriminative ERP information. To enhance task-relevant activity before deep feature learning, the proposed architecture incorporates xDAWN spatial filtering, which projects the original multichannel EEG into a compact spatial representation optimized for ERP enhancement.

Unlike purely data-driven spatial feature extraction, xDAWN explicitly exploits the reproducible temporal structure of ERP responses. The algorithm first estimates a class-specific ERP template by averaging the EEG trials belonging to each class. Let $X_i^{(k)} \in \mathbb{R}^{C\times T}$ denote the $i$-th EEG trial of class $k$, where $C = 64$ is the number of EEG channels and $T = 192$ is the number of temporal samples. The class-specific ERP template is computed as

$$P^{(k)} = \frac{1}{N_k}\sum_{i=1}^{N_k} X_i^{(k)} \tag{1}$$

where $N_k$ denotes the number of trials belonging to class $k$.

The spatial filters are subsequently estimated by maximizing the ratio between the energy of the class-specific ERP template and the overall EEG covariance. This optimization is formulated as the generalized Rayleigh quotient

$$V^* = \arg\max_{v} \frac{V^T P^{(k)} P^{(k)T} V}{V^T E E^T V} \tag{2}$$

where $X$ denotes the concatenation of EEG trials from all classes. Solving the corresponding generalized eigenvalue problem yields a set of class-specific spatial filters $V^{(k)} \in \mathbb{R}^{C \times a}$, where $n$is the number of retained filters for each class. The complete xDAWN projection matrix is then constructed by concatenating the filters from all classes, $\mathrm{V} = [V^{(1)}, V^{(2)}, \ldots, V^{(N)}] \in \mathbb{R}^{C \times M}$, where $N$ denotes the number of classes and $\boldsymbol{M} = \boldsymbol{a} \times \boldsymbol{b}$ represents the total number of retained spatial components. In the proposed binary P300 detection framework ($b = 2$), 16 spatial components are retained, resulting in a projection matrix $V \in \mathbb{R}^{64 \times 16}$. The spatially filtered representation of the $i$-th EEG trial is then obtained as

$$X_{\text{filtered}}^{(i)} = V^T X_i \in \mathbb{R}^{16 \times 192} \tag{3}$$

Thus, the original 64-channel EEG is transformed into 16 virtual channels, each representing an optimized linear combination of the original electrodes that emphasizes P300-related activity while reducing spatial redundancy. Unlike conventional xDAWN pipelines, these virtual channels directly serve as the input to the subsequent filter-bank and spatio-temporal convolutional stages, preserving ERP-oriented spatial priors within the end-to-end learning framework. Finally, each projected EEG segment is standardized using Z-score normalization,

$$\hat{X}_{\text{filtered}}^{(i)} = \frac{X_{\text{filtered}}^{(i)} - \mu}{\sigma} \tag{4}$$

where $\mu$ and $\sigma$ denote the mean and standard deviation computed independently for each projected channel. This normalization ensures consistent feature scaling across trials before the subsequent frequency-selective and spatio-temporal representation learning stages.

**Filter-Bank Convolution Layer:** A filter-bank convolution layer comprising six temporal convolutional filters with kernel size ((48,1)) is subsequently applied to the spatially filtered representation. The convolution operates along the temporal dimension while preserving the spatial-channel dimension. Each filter has a temporal extent of 48 samples, corresponding to 200 ms at the 240-Hz sampling rate. The six filters are initialized using finite impulse response (FIR) band-pass filters designed with a Kaiser-window method to represent six conventional EEG frequency bands: delta (1–4 Hz), theta (4–8 Hz), alpha (8–13 Hz), beta-1 (13–20 Hz), beta-2 (20–30 Hz), and gamma (30–40 Hz). The filter coefficients are kept fixed during optimization to ensure that each convolutional kernel consistently represents its predefined frequency band. Consequently, the contribution of individual frequency bands can also be examined independently by selectively applying the corresponding filter. For example, initializing a single ((48,1)) kernel using a 13–20 Hz FIR band-pass filter enables the network to perform classification exclusively using beta-2 activity.

**EEG Spatial Channel Optimization Layer:** For each of the six frequency bands, the network subsequently learns optimal linear combinations of the spatial EEG components. This operation is implemented using a depthwise convolution with a kernel size of ((1,16)), which operates across the 16 spatial components generated by xDAWN independently within each frequency band. The number of filters assigned to each frequency band is controlled by the hyperparameter (D), referred to as the depth multiplier. Different values of (D) were evaluated empirically, and (D=11) was selected for the final architecture. With appropriate padding, the resulting representation has dimensions ((66,192,1)), corresponding to (6* depth multiplier =66) learned spatial-frequency feature maps. This operation enables the network to learn frequency-specific spatial projections while maintaining a compact representation.

**Batch Normalization and Activation:** The output of the preceding spatial convolution is processed by batch normalization followed by an exponential linear unit (ELU) activation function. These operations preserve the dimensionality of the representation while normalizing intermediate activations and introducing nonlinear transformations. No additional spatial or temporal dimensions are introduced at this stage.

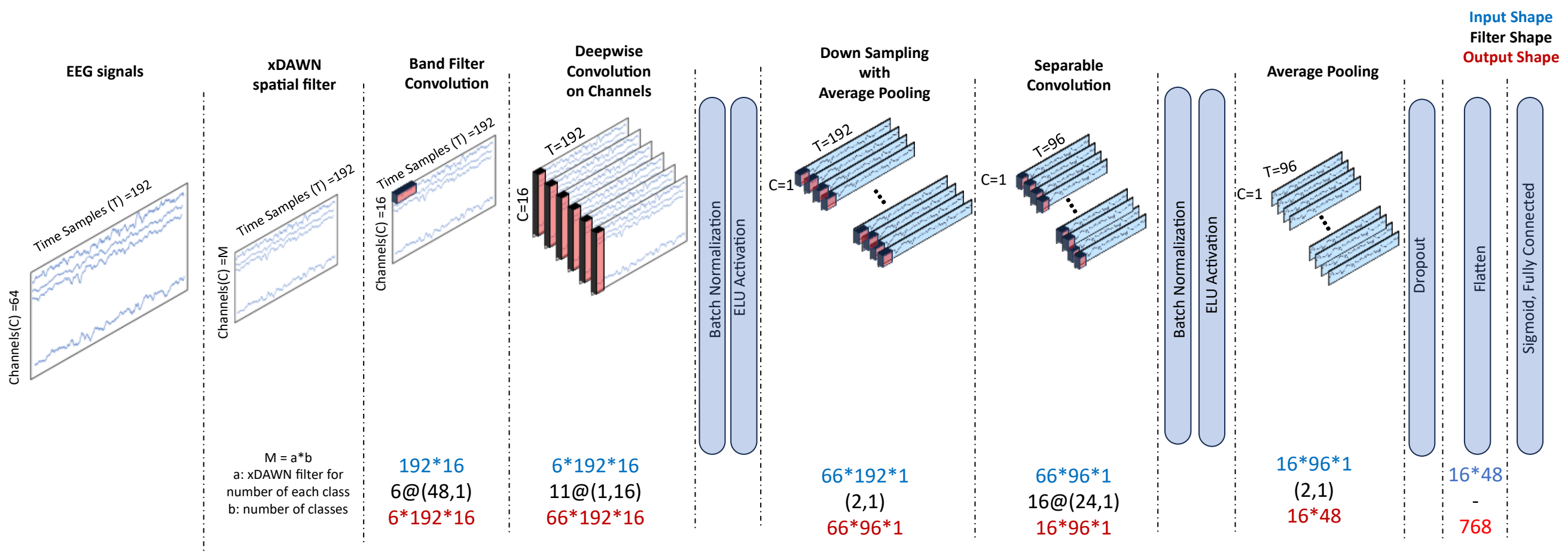

*Figure 2. Overview of our proposed architecture for P300 detection from EEG signals.*

**Temporal Down-sampling Layer:** To reduce the temporal dimensionality of the feature representation, average pooling with a kernel size of ((2,1)) is applied along the temporal dimension. This operation reduces the number of temporal samples from 192 to 96, resulting in an output tensor of dimensions ((66,96,1)).

**Temporal Filtering Layer (Depthwise-Separable Convolution):** To construct richer temporal feature representations by integrating information across neighboring temporal samples, depthwise-separable two-dimensional convolutions with a kernel size of ((24,1)) are employed. The number of separable convolutional filters is treated as a network hyperparameter and was set to 16 in the final architecture. With appropriate padding, this layer produces an output tensor of dimensions ((16,96,1)). The depthwise-separable convolution consists of two sequential operations. First, a depthwise convolution with a kernel size of ((24,1)) extracts temporal features independently from the input feature maps. This operation is followed by 16 pointwise convolutions with kernel size ((1,1)), which linearly combines the resulting feature maps across channels. This factorized formulation substantially reduces the number of trainable parameters compared with a conventional convolution, while preserving the ability to perform channel-wise temporal feature extraction before pointwise feature integration.

**Temporal Down-sampling Layer:** A second average-pooling operation with a kernel size of ((2,1)) is applied along the temporal dimension to further reduce the representation size. This operation decreases the temporal dimension from 96 to 48 samples, producing an output tensor of dimensions ((16,48,1)).

Flatten Layer: The resulting four-dimensional feature tensor is flattened into a one-dimensional feature vector containing (16*48=768) elements.

**Output Layer:** The final classification stage consists of a fully connected (dense) layer with a sigmoid activation function. The layer maps the 768-dimensional feature representation to a single probability corresponding to the binary classification of the input EEG trial as target or non-target.

### *2.3. Training and Optimization*

The proposed network was implemented in PyTorch [24] and trained on an NVIDIA GeForce GTX 1080 GPU. Also, xDAWN spatial filters were estimated using the pyRiemann package [25]. The network was optimized using the Adam optimizer [26] with an initial learning rate of 0.001 for 100 epochs, using a mini-batch size of 64 and a dropout rate of 0.2 to reduce overfitting. All convolutional layers were initialized using a zero-mean Gaussian distribution, except for the filter-bank convolution layer. Instead of random initialization, this layer was initialized with Kaiser-window FIR band-pass filters corresponding to six predefined EEG frequency bands and remained fixed throughout training. This design preserves physiologically meaningful spectral representations while avoiding unnecessary optimization of the filter-bank coefficients. Given the approximately 5:1 imbalance between non-target and target trials in the P300 speller datasets, we employed Focal Loss [27] instead of conventional binary cross-entropy to place greater emphasis on difficult-to-classify samples. The loss function is defined as

$$L_{FL} = -\sum_{k=1}^{N} w_k\, t_k (1 - p_k)^{\gamma} \log(p_k) \tag{5}$$

where $p_k$ denotes the predicted probability of class $k$, $t_k$ is the corresponding ground-truth label, $w_k$ is the class-specific weighting factor, and $\gamma$ controls the degree to which well-classified samples are down-weighted. Following the original formulation of Focal Loss, $\gamma = 2$ was used in all experiments [27].

## 3. RESULTS

### *3.1. Evaluation Metric for P300 Speller Performance*

The primary performance metric for evaluating the proposed P300 speller is the Information Transfer Rate (ITR) [9], which jointly captures decoding accuracy and communication speed and is therefore more representative of practical BCI performance than classification accuracy alone. ITR is reported in bits per minute and quantifies the effective rate at which information can be transmitted from the user to the external device. For a P300 speller with possible characters, character-level accuracy, and average time required to select one character, ITR is calculated as [20], [21]:

$$ITR = \left(\left(\log_2 N + P\log_2 P + (1-P)\log_2 \frac{1-P}{N-1}\right) \times 60\right) / T \quad (6)$$

In the present experiments, N=36, corresponding to the character matrix. The character-selection time T depends on the number of stimulus repetitions $Nr$ and the timing parameters of the P300 paradigm and is calculated as:

$$T = 2.5 + \left((0.100s + 0.075s) \times 12\right) \times Nr; \quad (7)$$
$$1 \le Nr \le 15$$

where $Nr \in \{1, \ldots, 15\}$ denotes the number of repetitions, 12 is the number of row/column flashes required per repetition, each flash cycle consists of a 100-ms illumination followed by a 75-ms inter-stimulus interval, and 2.5 s represents the inter-character interval. Accordingly, ITR is evaluated across different numbers of stimulus repetitions to characterize the trade-off between character recognition accuracy and communication speed.

### *3.2. Baseline Methods*

To comprehensively evaluate the performance of the proposed method, we benchmarked it against representative and established baseline methods spanning conventional machine-learning and signal-processing approaches, deep convolutional networks, capsule-based architectures, and hybrid deep learning–Riemannian methods. Specifically, the baselines include ESVM, DCPM, MCNN-1, BN3, DeepConvNet, ShallowConvNet, EEGNet, CNN-RG-MINMAX, ERP-CapsNet, and ST-CapsNet. Detailed descriptions of these methods are provided below.

ESVM employs an ensemble of SVM classifiers for P300 target/non-target discrimination, using class-balancing strategies to address the inherent class imbalance and improve ensemble robustness and classification performance [28].

**MCNN-1** is an ensemble of five CNN-1 networks, each comprising spatial and temporal convolutional layers followed by fully connected layers, trained on class-balanced data partitions for binary P300 classification [19]. **BN3** extends CNN-based P300 detection by incorporating batch normalization into the network to improve training stability and mitigate overfitting [21], while **DeepConvNet** employs a deeper sequence of convolutional blocks to learn hierarchical EEG representations [18]. **ShallowConvNet**, in contrast, is explicitly inspired by the filter-bank common spatial pattern (FBCSP) framework and uses temporal and spatial convolutions followed by nonlinear squaring and mean pooling to learn representations related to band-power features [18]. **EEGNet** provides a compact alternative based on depthwise and depthwise-separable convolutions, substantially reducing the parameter count while retaining explicit temporal and spatial feature extraction [17].

More recent approaches further incorporate structured representations and advanced feature-learning mechanisms. **DCPM** is a discriminative spatial-temporal pattern-matching method that learns class-specific canonical ERP templates and discriminative spatial filters, then classifies single-trial ERPs based on their matching to the learned canonical patterns [29]. **CNN-RG-MINMAX** is a hybrid P300 decoder that maps EEG covariance matrices to a discriminative low-dimensional SPD representation using MINMAX Riemannian mapping, followed by CNN-based classification [20].

**ERP-CapsNet** introduces one-dimensional convolutional capsule networks for P300 detection, using convolutional feature extraction followed by capsule-based representation learning to preserve discriminative feature relationships [22]. **ST-CapsNet** extends this framework by incorporating dedicated spatial and temporal attention modules before capsule-based feature extraction, enabling the model to selectively emphasize event-related spatial and temporal information in the EEG [23].

Collectively, these baselines span complementary modeling paradigms, ranging from ensemble-based machine learning and discriminative pattern matching to compact and deep CNNs, Riemannian hybrid models, and attention-enhanced capsule networks. This selection enables evaluation of the proposed architecture against methods that capture complementary aspects of EEG structure, including spatial, temporal, spectral, geometric, and hierarchical representations.

### *3.3. Performance of Symbol Recognition*

The symbol-level recognition results are summarized in Tables 3 and 4 for BCI Competition II and BCI Competition III, respectively. On BCI Competition II (Table 3), most methods eventually reached near-perfect recognition performance as the number of repetitions increased; however, notable differences were observed in the low-repetition regime, which is of greater practical importance for maximizing communication speed. The proposed method achieved full symbol recognition after only four repetitions, demonstrating a convergence rate comparable to the strongest deep-learning baselines while substantially outperforming conventional approaches such as ESVM and DCPM during the early repetitions. A similar trend was observed on BCI Competition III (Table 4), where the proposed model consistently ranked among the best-performing methods across repetitions and achieved the highest average recognition performance in the later stages. In particular, the performance improved steadily from 34 correctly recognized symbols after the first repetition to 98 symbols after 15 repetitions, indicating stable accumulation of discriminative evidence across successive trials. Collectively, these results suggest that the proposed architecture not only maintains high final recognition accuracy but also enables rapid and reliable symbol identification with relatively few repetitions, a property that is particularly desirable for practical P300-speller systems due to its direct impact on information transfer rate and user interaction efficiency.

#### *3.3.1. Performance for BCI Competition II*

Figure. 3 compares the proposed model with state-of-the-art methods using ITR on the BCI Competition II (Dataset IIb). Because ITR jointly reflects character-selection accuracy and communication speed, it is one of the most informative metrics for evaluating the practical efficiency of P300 speller systems.

As shown in Figure. 3, all methods exhibit a similar overall trend: ITR increases during the initial repetitions, reaches a maximum operating point, and subsequently declines as additional repetitions are introduced. This behavior reflects the well-established **speed–accuracy trade-off** in P300-based BCIs. Although increasing the number of repetitions generally improves classification accuracy through evidence accumulation, these gains gradually saturate, whereas the selection time increases linearly. Consequently, beyond an optimal repetition number, the additional time required outweighs the marginal improvement in accuracy, resulting in a reduction in ITR. The early repetition range (1–4 repetitions) is particularly important because it directly determines communication efficiency. Within this operating region, the proposed model consistently achieves the strongest performance, reaching its maximum ITR at 2–3 repetitions. This behavior suggests that the proposed architecture is able to extract discriminative P300 representations with relatively few stimulus repetitions, thereby reducing the amount of evidence required before reliable decoding can be achieved.

Compared with the proposed approach, ESVM attains a relatively high ITR at the first repetition but exhibits a more rapid decline as repetitions increase. Deep learning baselines such as DeepConvNet and BN3 remain competitive across the repetition range, yet their peak ITR remains below that of the proposed model. In contrast, methods including ShallowConvNet and DCPM reach their highest performance at larger repetition numbers, indicating a less favorable balance between decoding speed and communication efficiency under this evaluation protocol.

Overall, the ITR analysis indicates that the proposed model achieves its most effective operating point at 2–3 repetitions, where it provides the highest communication efficiency while maintaining strong decoding performance. Rather than requiring additional stimulus repetitions to approach peak performance, the model reaches its optimal operating region early, supporting its potential for fast and computationally efficient P300 speller decoding in practical BCI applications.

#### *3.3.2. Performance for BCI Competition III (Subjects A & B)*

To further evaluate the robustness of the proposed model under a more challenging decoding scenario, its performance was assessed on BCI Competition III (Dataset II) using the average results across Subjects A and B. Compared with BCI Competition II, this dataset presents a more demanding evaluation setting because of its lower signal-to-noise ratio and greater inter-subject variability, making it a more stringent benchmark for assessing the ability of a decoder to extract discriminative P300 representations from noisy EEG signals. The ITR results for all compared methods are presented in Figure. 4. Compared with the BCI Competition II results, all methods exhibit lower peak ITR values on this dataset, with the highest performance remaining around 18 bits/min. This overall reduction reflects the increased difficulty of the decoding task under the BCI Competition III evaluation protocol. Nevertheless, the same speed–accuracy trade-off observed on the previous dataset remains evident: as repetitions increase, the gain in decoding accuracy gradually diminishes while the communication time continues to increase, causing the ITR to decline beyond each model's optimal operating point. For many baseline methods, this peak shifts toward 3–4 repetitions, indicating a greater reliance on repeated stimulus presentations before reaching their maximum communication efficiency. Within this more challenging operating regime, the proposed model achieves the strongest performance, reaching its maximum ITR of approximately 17.8 bits/min at two repetitions. Unlike most competing approaches, whose peak performance occurs at later repetitions, the proposed architecture reaches its optimal operating point with substantially fewer stimulus repetitions. This behavior suggests that the combination of xDAWN-based spatial enhancement and hierarchical spatio-temporal representation learning enables the network to recover discriminative P300 information efficiently, even under more challenging recording conditions. Among the baseline methods, DeepConvNet provides the closest performance to the proposed approach across the repetition range, although its peak ITR remains consistently below that of the proposed model. Most of the remaining methods cluster around intermediate performance levels and require additional repetitions to approach their maximum ITR, resulting in lower communication efficiency under this evaluation protocol. In contrast, DCPM exhibits comparatively lower ITR throughout the repetition range, indicating reduced decoding efficiency on this dataset. Overall, the results on BCI Competition III reinforce the observations obtained from the previous dataset. Despite the increased decoding difficulty, the proposed model preserves its ability to reach peak communication efficiency at an early repetition stage. Consequently, two repetitions represent the optimal operating point for the proposed architecture under this evaluation protocol, providing the highest ITR while maintaining strong decoding performance. These findings support the potential of the proposed architecture as a compact and efficient decoder for rapid P300 speller communication under challenging EEG recording conditions.

### *3.4. Computational Efficiency Analysis*

Practical deployment of a P300 speller requires more than high decoding accuracy. An effective decoder must simultaneously minimize the amount of neural evidence required for reliable decisions while maintaining a computational complexity compatible with low-latency processing. Accordingly, the proposed architecture is evaluated from two complementary perspectives: **information efficiency**, which reflects how many stimulus repetitions are required before reaching an effective operating point, and **computational efficiency**, which reflects the complexity of the network required to process those EEG signals.

#### *3.4.1. Information Efficiency*

The ITR analyses on both BCI Competition II and BCI Competition III consistently show that the proposed model reaches its most effective operating region after only two repetitions, while maintaining competitive performance at the third repetition. We therefore adopt two repetitions as the practical operating configuration, since it provides the highest communication efficiency while minimizing the amount of neural evidence required before reliable decoding can be achieved.

Based on the timing model defined in Eq. (7), where each repetition of the standard 6×6 P300 speller consists of 12 stimulus flashes (six rows and six columns), a two-repetition configuration corresponds to a total character-selection time of approximately 6.7 s, requiring only 24 stimulus presentations. This operating point represents a favorable trade-off between decoding reliability and communication speed, as reducing the required number of repetitions directly shortens character-selection time. Rather than relying on prolonged stimulus averaging, the proposed architecture extracts discriminative P300 representations early in the evidence accumulation process, enabling efficient decoding from a relatively small amount of EEG evidence.

#### *3.4.2. Computational Efficiency*

Information efficiency alone is insufficient for practical P300-speller deployment; the decoder must also remain computationally lightweight for resource-constrained and low-latency BCI systems. Because inference latency depends on hardware and implementation details and was not directly measured in this study, computational efficiency is assessed using the number of trainable parameters as a hardware-independent measure of model complexity.

Table 2 compares the proposed architecture with representative end-to-end CNN baselines that span the principal performance–complexity trade-off observed in the ITR analysis, namely DeepConvNet, ShallowConvNet, BN3, and EEGNet. These models were selected because they represent architecturally comparable convolutional P300 decoders while covering complementary operating points ranging from high-capacity CNNs to compact lightweight designs. In contrast, conventional machine-learning methods (ESVM and DCPM) and hybrid or capsule-based approaches (MCNN-1, CNN-RG-MINMAX, ERP-CapsNet, and ST-CapsNet) employ substantially different computational mechanisms, making parameter-count comparisons less informative than comparisons within a common end-to-end CNN framework. The proposed architecture contains 5,100 trainable parameters, representing a substantial reduction compared with larger CNN architectures such as

*Table2 .Number of trainable parameters of the proposed model and representative end-to-end CNN-based P300 decoders.*

| Method | Parameters |
|---|---|
| ShallowConvNet [18] | 105,282 |
| DeepConvNet [18] | 176,927 |
| BN3 [21] | 47,681 |
| EEGNet [17] | 2,354 |
| Our Method | 5,100 |

*Table 3. Number of correctly recognized symbols per repetition for all compared models on BCI Competition II (Dataset IIb).*

| Model | Repetition | | | | | | | | | | | | | |
|---|---|---|---|---|---|---|---|---|---|---|---|---|---|---|
| | 1 | 2 | 3 | 4 | 5 | 6 | 7 | 8 | 10 | 11 | 12 | 13 | 14 | 15 |
| ESVMs [28] | 25 | 29 | 28 | 30 | 31 | 31 | 31 | 31 | 31 | 31 | 31 | 31 | 31 | 31 |
| DeepConvNet [18] | 24 | 29 | 31 | 31 | 31 | 31 | 31 | 31 | 31 | 31 | 31 | 31 | 31 | 31 |
| ShallowConvNet [18] | 23 | 26 | 31 | 31 | 31 | 31 | 31 | 31 | 31 | 31 | 31 | 31 | 31 | 31 |
| BN3 [21] | 24 | 23 | 27 | 28 | 29 | 30 | 30 | 30 | 31 | 31 | 31 | 31 | 31 | 31 |
| MCNN-1[19] | 23 | 26 | 29 | 31 | 31 | 31 | 31 | 31 | 31 | 31 | 31 | 31 | 31 | 31 |
| CNN-RG-MINMA [20] | 22 | 24 | 30 | 30 | 31 | 31 | 31 | 31 | 31 | 31 | 31 | 31 | 31 | 31 |
| ERP-CapsNet [22] | 24 | 27 | 29 | 31 | 31 | 31 | 31 | 31 | 31 | 31 | 31 | 31 | 31 | 31 |
| ST-CapsNet [23] | 27 | 29 | 30 | 31 | 31 | 31 | 31 | 31 | 31 | 31 | 31 | 31 | 31 | 31 |
| DCPM [29] | 18 | 22 | 27 | 29 | 28 | 30 | 31 | 31 | 31 | 31 | 31 | 31 | 31 | 31 |
| EEGNet [17] | 22 | 28 | 31 | 31 | 31 | 31 | 31 | 31 | 31 | 31 | 31 | 31 | 31 | 31 |
| Our Method | 23 | 29 | 30 | 31 | 31 | 31 | 31 | 31 | 31 | 31 | 31 | 31 | 31 | 31 |

*Table 4. Average number of correctly recognized symbols per repetition across Subjects A and B from BCI Competition III for all compared models.*

| Model | Subject (A, B) | Repetition | | | | | | | | | | | | | | |
|---|---|---|---|---|---|---|---|---|---|---|---|---|---|---|---|---|
| | | 1 | 2 | 3 | 4 | 5 | 6 | 7 | 8 | 9 | 10 | 11 | 12 | 13 | 14 | 15 |
| ESVMs [28] | Mean | 29.5 | 48 | 62.5 | 67.5 | 74.5 | 81.5 | 80.5 | 84.5 | 89.5 | 92 | 93.5 | 95.5 | 96 | 97 | 98 |
| DeepConvNet[18] | Mean | 32 | 49 | 63 | 69.5 | 75.5 | 83 | 89 | 90 | 88.5 | 93 | 93 | 89 | 95.5 | 97 | 97.5 |
| ShallowConvNet [18] | Mean | 32.5 | 40.5 | 52.5 | 56.5 | 62 | 70.5 | 75 | 75.5 | 80.5 | 83 | 88 | 93.5 | 89 | 90.5 | 92.5 |
| BN3 [21] | Mean | 34.5 | 49 | 64 | 70 | 74.5 | 78.5 | 81.5 | 86 | 88 | 90.5 | 92 | 91.25 | 94 | 95 | 96.5 |
| MCNN-1[19] | Mean | 33.5 | 44.75 | 58.25 | 63.25 | 68.25 | 74.5 | 78.25 | 80.75 | 84.25 | 86.75 | 90 | 94 | 91.5 | 92.75 | 94.5 |
| CNN-RG-MINMA [20] | Mean | 31 | 45 | 57 | 66 | 71.5 | 73.5 | 80 | 82 | 85 | 90.5 | 91 | 92.62 | 93.5 | 95 | 97 |
| ERP-CapsNet [22] | Mean | 32.25 | 44.87 | 57.62 | 64.62 | 69.87 | 74 | 79.12 | 81.37 | 84.62 | 88.62 | 90.5 | 92.5 | 92.5 | 93.87 | 95.75 |
| ST-CapsNet [23] | Mean | 28 | 44 | 58.5 | 64.5 | 73.5 | 76.5 | 82 | 85 | 86.5 | 89.5 | 90.5 | 88 | 91.5 | 94.5 | 95.5 |
| DCPM [29] | Mean | 22 | 36 | 46.5 | 54 | 60 | 66.5 | 73 | 77.5 | 82 | 85 | 88 | 95 | 88 | 91 | 93.5 |
| EEGNet [17] | Mean | 31 | 43.5 | 58.5 | 67.5 | 73 | 83 | 84.5 | 85 | 88.5 | 92 | 93 | 96 | 96 | 95.5 | 96.5 |
| Our Method | Mean | 34 | 57 | 64.5 | 70.5 | 76.5 | 82 | 84 | 87.5 | 90.5 | 92.5 | 94.5 | 96.5 | 97 | 97 | 98 |

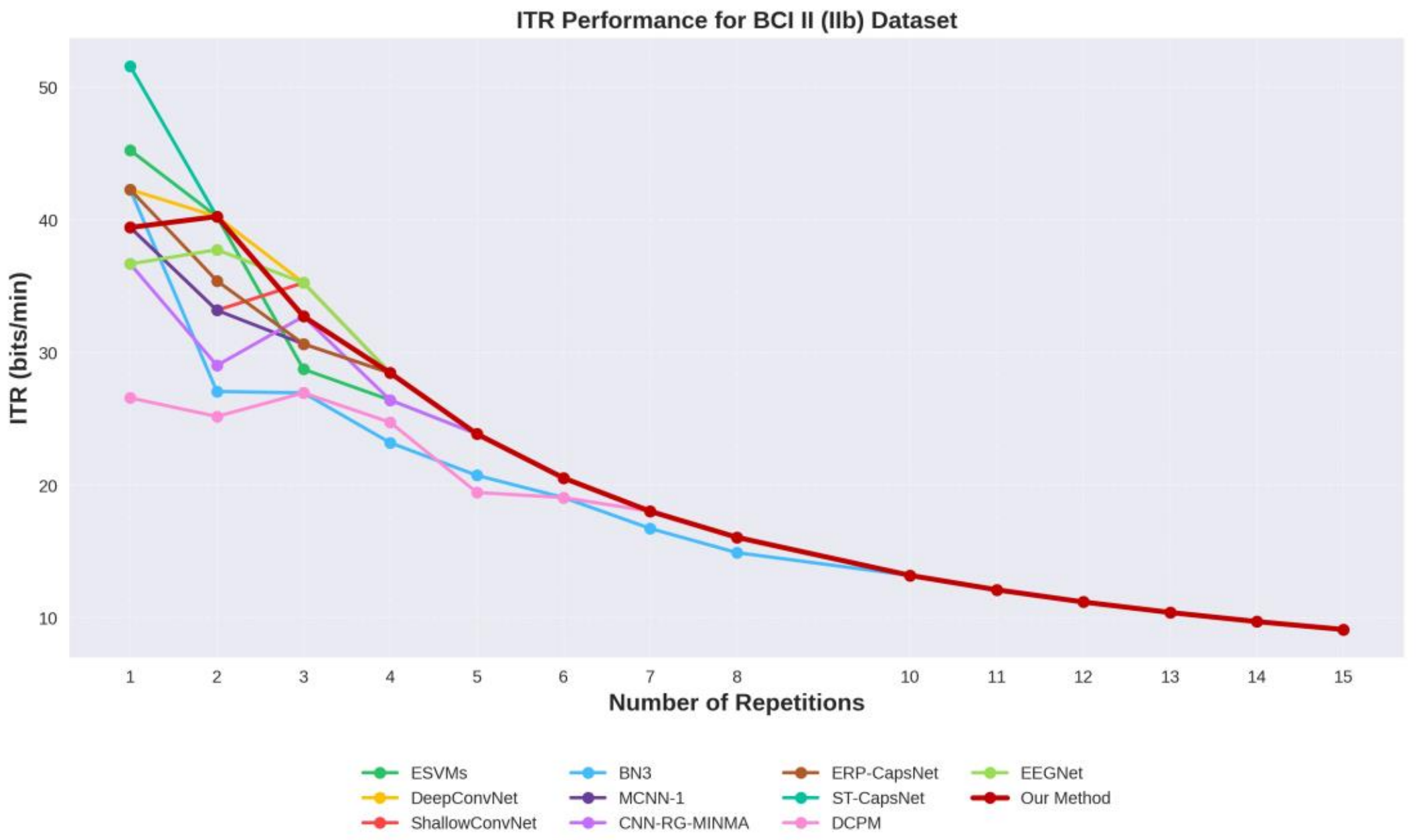

*Figure 3. ITR Comparison under 15 Repetitions between Our Method and Baseline Methods on BCI Competition II (Dataset IIb).*

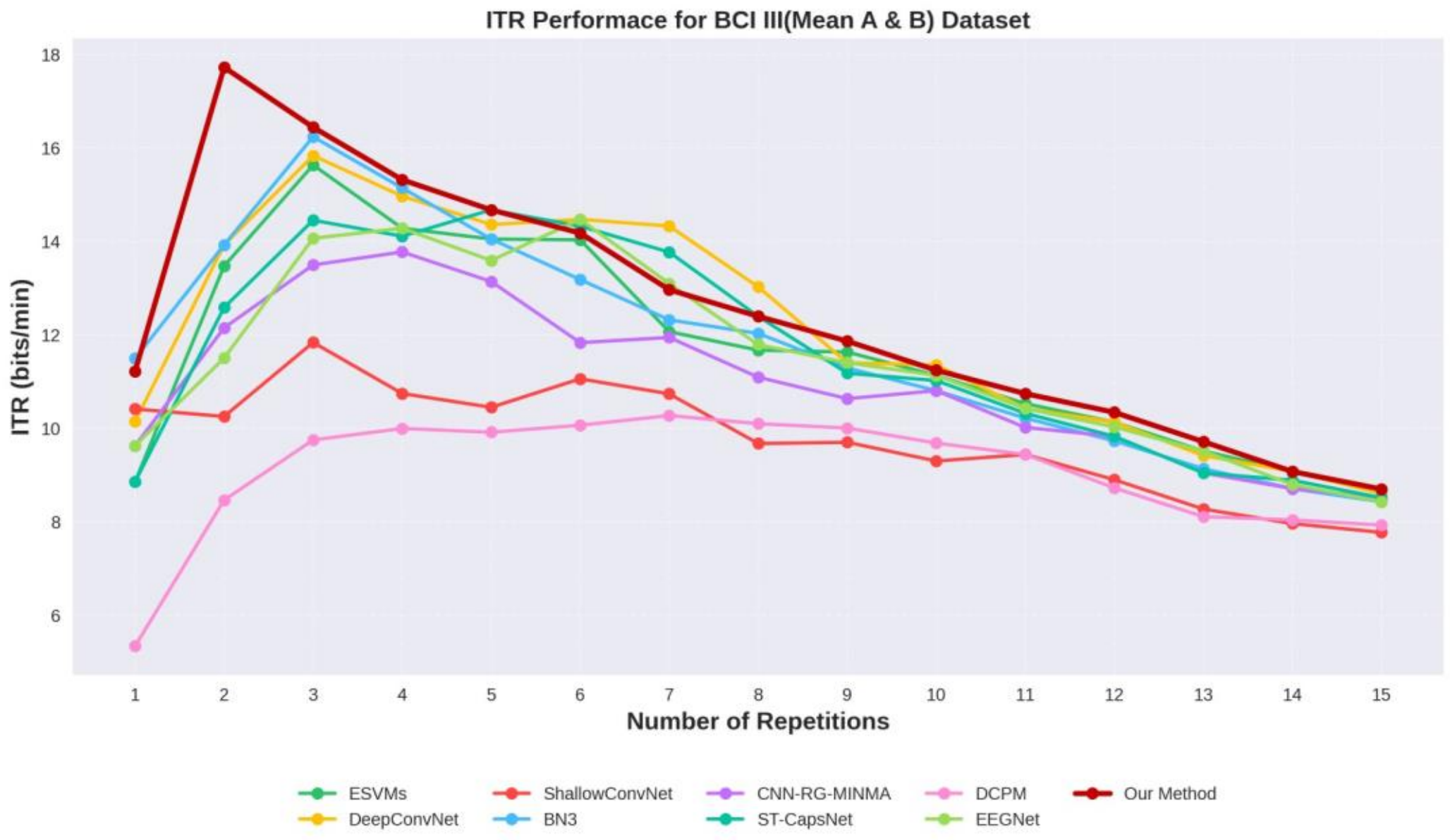

*Figure 4. ITR Comparison under 15 Repetitions between Our Method and Baseline Methods on BCI Competition III, Based on the Mean Performance of Two Subjects (A and B).*

DeepConvNet, ShallowConvNet, and BN3. Although EEGNet remains the smallest model among the compared CNNs with 2,354 parameters, the proposed network operates within the same compact-model regime while maintaining competitive decoding performance. The compact parameter budget is achieved through three complementary architectural decisions. First, xDAWN reduces the original 64-channel EEG to a compact 16-component spatial representation before deep feature learning, substantially decreasing the dimensionality of subsequent processing. Second, the initial FIR-based filter-bank convolution performs frequency-specific temporal decomposition using fixed kernels, eliminating the need to learn six independent temporal filters while preserving predefined EEG-band representations. Third, the combination of a depthwise spatial convolution across the xDAWN components and a subsequent depthwise-separable temporal convolution enables efficient frequency-specific spatial optimization and temporal feature modeling with considerably fewer parameters than conventional convolutional layers. Taken together with the information-efficiency analysis, these results indicate that the proposed architecture occupies a favorable position on the performance–efficiency frontier. Rather than relying on a high-capacity network to achieve competitive decoding performance, it combines a compact parameter footprint with an effective operating point at only two repetitions, supporting practical and resource-efficient P300-speller deployment under the evaluated benchmark protocol.

## 4. DISCUSSION

The results indicate that our proposed method provides a favorable balance between decoding accuracy, repetition efficiency, and model compactness, which are particularly important for practical P300 spellers. Across the evaluated repetition levels, the proposed model reaches its highest information-transfer efficiency at substantially fewer repetitions than most competing approaches. On BCI Competition II, the optimal operating region occurs at approximately two to three repetitions, while the same two-repetition operating point is maintained on the more challenging BCI Competition III data. This behavior is important because increasing the number of repetitions can improve confidence in P300 detection but simultaneously increases the communication time per character. Thus, the advantage of the proposed model is not simply reflected in classification accuracy, but in its ability to extract sufficient discriminative information from fewer stimulus repetitions.

The robustness of this behavior on BCI Competition III is particularly notable. This dataset provides a more demanding evaluation setting because of lower SNR and inter-subject variability, yet the proposed architecture retains high performance and reaches its optimal operating point with only two repetitions. The consistent behavior across Subjects A and B suggests that explicitly incorporating spatial ERP enhancement through xDAWN before deep feature learning may help the network extract P300-related structure from noisy multichannel EEG without requiring a large unconstrained deep architecture. This observation is consistent with the design rationale of the model: rather than forcing the CNN to learn all spatial structure directly from the raw EEG, xDAWN provides an ERP-oriented spatial representation, after which the convolutional layers focus on frequency-specific and temporal discrimination.

The ITR analysis further demonstrates that the principal benefit of the proposed architecture lies in the **speed–accuracy trade-off**. While the compared methods generally exhibit an increase in performance with additional repetitions followed by a decline in ITR, the proposed model reaches its maximum information-transfer efficiency at an early stage. The reported ITR of 40.5 bits/min on BCI Competition II and 17.8 bits/min on BCI Competition III therefore indicate that the architecture can achieve a useful communication rate without relying on extensive stimulus averaging. This is particularly relevant for P300 spellers, where reducing the number of required repetitions directly reduces interaction time.

An additional advantage is the computational compactness of the proposed network. xDAWN first reduces the original 64-channel representation to 16 spatial components, while the subsequent depthwise and depthwise-separable convolutions limit the number of trainable parameters. The resulting model contains 5,100 trainable parameters, substantially fewer than DeepConvNet (176,927), ShallowConvNet (105,282), and BN3 (47,681), while remaining competitive with the highly compact EEGNet (2,354 parameters). This comparison suggests that the performance gain is not obtained simply by increasing model capacity. Instead, the results support the value of combining domain-informed spatial preprocessing with structured frequency and temporal representation learning.

Taken together, these findings suggest that the proposed architecture is particularly well suited to the constraints of P300-based BCI decoding, where EEG data are noisy, training data are limited, and communication speed is as important as classification accuracy. Nevertheless, the current evaluation is based on publicly available offline datasets and therefore does not by itself establish real-world online performance. The next step should be a subject-independent and online evaluation, including end-to-end latency, calibration requirements, and performance in users with substantially different P300 morphology. Such experiments would determine whether the observed reduction in repetition requirements translates directly into improved real-time BCI communication.

### *4.1. Limitations and future directions*

Despite the promising decoding performance of the proposed architecture, several directions remain open for further investigation. First, the present evaluation is based on a limited number of subjects, with one subject from BCI Competition II and two subjects from BCI Competition III. Although the model demonstrates consistent performance across the available participants, broader cross-subject and cross-session validation is required to establish its generalizability across inter-individual variations in P300 morphology, latency, and spatial distribution. Future studies should therefore investigate subject-independent learning and lightweight subject-specific adaptation to reduce calibration requirements.

Second, although xDAWN provides an effective ERP-informed spatial projection and substantially reduces the dimensionality of the multichannel EEG, its fixed projection may constrain the representation available to the subsequent deep-learning layers. A promising extension would therefore be a dual-path architecture, in which the xDAWN-enhanced representation is processed in parallel with a learnable spatial pathway operating directly on the EEG, followed by a compact feature-fusion mechanism. Such a design could retain the advantages of the established ERP prior while allowing the network to recover complementary spatial information not captured by the xDAWN projection.

Finally, the current temporal encoder relies primarily on a single-scale depthwise-separable convolution. While this design provides an effective compromise between temporal representation capacity and computational complexity, P300 responses can exhibit subject- and trial-dependent temporal variability. Future architectures could therefore incorporate multi-scale temporal convolutions with different kernel sizes or receptive fields, enabling the model to capture both short- and long-range temporal dynamics while preserving its compact parameterization. Together, these extensions could lead toward more adaptive, generalizable, and computationally efficient P300 decoders suitable for robust real-world BCI applications.

## 5. CONCLUSION

This study introduced a compact spatio-temporal deep learning network for P300 detection from EEG, combining established ERP-specific signal-processing priors with learnable representation learning. The architecture first employs xDAWN spatial filtering to enhance P300-related activity and reduce the spatial dimensionality, followed by a fixed FIR filter bank to preserve physiologically meaningful frequency-specific information. Frequency-specific spatial learning and depthwise-separable temporal convolutions then enable end-to-end discriminative feature extraction with only 5,100 trainable parameters.

Evaluated on publicly available P300 speller datasets from the BCI Competitions II and III, the proposed method consistently achieved competitive target/non-target decoding performance and higher information transfer rates with fewer stimulus repetitions than the evaluated state-of-the-art methods. In particular, it achieved an ITR of 17.8 bits/min at two repetitions on BCI Competition III and 40.5 bits/min on BCI Competition II, demonstrating the potential of the proposed compact architecture for efficient P300 speller decoding. Overall, these findings support the value of integrating domain-informed signal processing with parameter-efficient deep learning for accurate and rapid EEG-based BCI decoding.

### Author Contributions

M.M.H.B. contributed to Conceptualization, Methodology, Investigation, Visualization, Writing – Original Draft, and Writing – Review & Editing. M.R.D. contributed to Supervision and Writing – Review & Editing. All authors read and approved the final manuscript.

### Funding

The research conducted for this paper did not receive any financial support or resources from any external sources.

### Conflict of Interest

The authors declare no conflict of interest.

### Ethics Statement

Ethics approval was not needed in this study.

## REFERENCES

[1] J. R. Wolpaw, "Brain–computer interfaces," in *Handbook of clinical neurology*, Elsevier, 2013, pp. 67–74.

[2] T. Ball, M. Kern, I. Mutschler, A. Aertsen, and A. Schulze-bonhage, "NeuroImage Signal quality of simultaneously recorded invasive and non-invasive EEG," vol. 46, pp. 708–716, 2009, doi: 10.1016/j.neuroimage.2009.02.028.

[3] B. J. Edelman, S. Zhang, G. Schalk, P. Brunner, and M. Review, "Non-Invasive Brain-Computer Interfaces : State of the Art and Trends," vol. 18, pp. 26–49, 2025.

[4] M. M. H. Bina, S. Baghernezhad, M. R. Daliri, and M. H. Moradi, "Neural Digital Twins: Toward Next-Generation Brain-Computer Interfaces," *arXiv Prepr. arXiv2601.01539*, 2026.

[5] J. N. Mak *et al.*, "Optimizing the P300-based brain–computer interface: current status, limitations and future directions," *J. Neural Eng.*, vol. 8, no. 2, p. 25003, 2011.

[6] T. W. Picton, "The P300 wave of the human event-related potential," *J. Clin. Neurophysiol.*, vol. 9, no. 4, pp. 456–479, 1992.

[7] M. Norizadeh, A. Mohammad, and L. Zhan, "Engineering Applications of Artificial Intelligence A novel multiclass-based framework for P300 detection in BCI matrix speller : Temporal EEG patterns of non-target trials vary based on their position to previous target stimuli," vol. 123, no. January, 2023.

[8] W. Yang, S. Xiang, and B. Huang, "Decoding P300 as a shared neural mechanism for oddball target detection and working memory updating," *Iscience*, vol. 29, no. 2, 2026.

[9] L. A. Farwell and E. Donchin, "Talking off the top of your head: toward a mental prosthesis utilizing event-related brain potentials," *Electroencephalogr. Clin. Neurophysiol.*, vol. 70, no. 6, pp. 510–523, 1988.

[10] A. Rakotomamonjy and V. Guigue, "BCI competition III: dataset II-ensemble of SVMs for BCI P300 speller," *IEEE Trans. Biomed. Eng.*, vol. 55, no. 3, pp. 1147–1154, 2008.

[11] B. Rivet, A. Souloumiac, V. Attina, and G. Gibert, "xDAWN Algorithm to Enhance Evoked Potentials: Application to Brain-Computer Interface," *IEEE Trans. Biomed. Eng.*, vol. 56, no. 8, pp. 2035–2043, 2009, doi: 10.1109/TBME.2009.2012869.

[12] H. Cecotti, M. P. Eckstein, and B. Giesbrecht, "Single-trial classification of event-related potentials in rapid serial visual presentation tasks using supervised spatial filtering," *IEEE Trans. neural networks Learn. Syst.*, vol. 25, no. 11, pp. 2030–2042, 2014.

[13] A. Barachant, S. Bonnet, M. Congedo, and C. Jutten, "Multiclass brain–computer interface classification by Riemannian geometry," *IEEE Trans. Biomed. Eng.*, vol. 59, no. 4, pp. 920–928, 2011.

[14] R. J. Kobler, J. Hirayama, Q. Zhao, and M. Kawanabe, "SPD domain-specific batch normalization to crack interpretable

unsupervised domain adaptation in,” no. NeurIPS, pp. 1–22, 2022.

[15] B. Boloukian and F. Safi-Esfahani, “Recognition of words from brain-generated signals of speech-impaired people: Application of autoencoders as a neural Turing machine controller in deep neural networks.,” *Neural networks Off. J. Int. Neural Netw. Soc.*, vol. 121, pp. 186–207, Jan. 2020, doi: 10.1016/j.neunet.2019.07.012.

[16] M. Eder, J. Xu, and M. Grosse-wentrup, “Journal of Neural Engineering OPEN ACCESS Benchmarking brain – computer interface algorithms : Riemannian approaches vs convolutional neural networks,” *J. Neural Eng.*, vol. 21, no. 4, p. 44002, 2024, doi: 10.1088/1741-2552/ad6793.

[17] V. J. Lawhern, A. J. Solon, N. R. Waytowich, S. M. Gordon, C. P. Hung, and B. J. Lance, “EEGNet: A compact convolutional neural network for EEG-based brain-computer interfaces,” *J. Neural Eng.*, vol. 15, no. 5, pp. 1–30, 2018, doi: 10.1088/1741-2552/aace8c.

[18] R. T. Schirrmeister *et al.*, “Deep Learning With Convolutional Neural Networks for EEG Decoding and Visualization,” vol. 5420, no. August, pp. 5391–5420, 2017, doi: 10.1002/hbm.23730.

[19] H. Cecotti and A. Gräser, “Convolutional neural networks for P300 detection with application to brain-computer interfaces,” *IEEE Trans. Pattern Anal. Mach. Intell.*, vol. 33, no. 3, pp. 433–445, 2011, doi: 10.1109/TPAMI.2010.125.

[20] S. N. Aghili and A. Erfanian, “A P300-based speller design using a MINMAX Riemannian geometry scheme and convolutional neural network,” *Ieee Access*, vol. 11, pp. 98633–98652, 2023.

[21] M. Liu, W. Wu, Z. Gu, Z. Yu, F. F. Qi, and Y. Li, “Deep learning based on Batch Normalization for P300 signal detection,” *Neurocomputing*, vol. 275, pp. 288–297, 2018, doi: 10.1016/j.neucom.2017.08.039.

[22] R. Ma, T. Yu, X. Zhong, Z. L. Yu, Y. Li, and Z. Gu, “Capsule Network for ERP Detection in Brain-Computer Interface,” *IEEE Trans. Neural Syst. Rehabil. Eng.*, vol. 29, pp. 718–730, 2021, doi: 10.1109/TNSRE.2021.3070327.

[23] Z. Wang, C. Chen, J. Li, F. Wan, Y. Sun, and H. Wang, “ST-CapsNet: Linking Spatial and Temporal Attention With Capsule Network for P300 Detection Improvement,” *IEEE Trans. Neural Syst. Rehabil. Eng.*, vol. 31, no. January, pp. 991–1000, 2023, doi: 10.1109/TNSRE.2023.3237319.

[24] A. Paszke *et al.*, “Pytorch: An imperative style, high-performance deep learning library,” *Adv. Neural Inf. Process. Syst.*, vol. 32, 2019.

[25] A. Barachant *et al.*, “pyRiemann/pyRiemann: V0. 10 (Version v0. 10)[Computer software],” *Zenodo*, 2026.

[26] D. P. Kingma and J. L. Ba, “Adam: A method for stochastic optimization,” *3rd Int. Conf. Learn. Represent. ICLR 2015 - Conf. Track Proc.*, pp. 1–15, 2015.

[27] T.-Y. Lin, P. Goyal, R. Girshick, K. He, and P. Dollár, “Focal loss for dense object detection,” in *Proceedings of the IEEE international conference on computer vision*, 2017, pp. 2980–2988.

[28] S. Kundu and S. Ari, “P300 detection using ensemble of SVM for brain-computer interface application,” in *2018 9th International Conference on Computing, Communication and Networking Technologies (ICCCNT)*, IEEE, 2018, pp. 1–5.

[29] X. Xiao, M. Xu, and J. Jin, “Discriminative Canonical Pattern Matching for Single-Trial Classification of ERP Components,” *IEEE Trans. Biomed. Eng.*, vol. 67, no. 8, pp. 2266–2275, 2020, doi: 10.1109/TBME.2019.2958641.